\documentclass[12pt]{article}
\usepackage{epsfig,amssymb,amsmath,psfrag,subfigure}

\newcommand{\insertfig}[2]{\mbox{\epsfxsize=#1cm \epsfbox{#2.eps}}}
\def \be  {\begin{equation}}
\def \ee  {\end{equation}}
\def \ba  {\begin{eqnarray}}
\def \ea  {\end{eqnarray}}
\def \baa {\begin{eqnarray*}}
\def \eaa {\end{eqnarray*}}
\def \bb  {\begin {thebibliography} }
\def \eb  {\end{thebibliography}}
\def \lab #1 {\label{#1}}

\newcommand\re[1]{(\ref{#1})}

\def \matrix #1 {\left(\begin{array}{cc} #1 \end{array}\right)}

\def \tr {\mathop{\rm tr}\nolimits}

\newcommand \vev [1] {\langle{#1}\rangle}

\newcommand{\bit}[1]{\mbox{\boldmath$#1$}}
\newcommand{\ft}[2]{{\textstyle\frac{#1}{#2}}}
\begin{document}

\begin{titlepage}

\thispagestyle{empty}

\vspace*{1cm}

\centerline{\large \bf OPE for null Wilson loops and open spin chains}

\vspace{1cm}

\centerline{\sc A.V. Belitsky}

\vspace{10mm}

\centerline{\it Department of Physics, Arizona State University}
\centerline{\it Tempe, AZ 85287-1504, USA}

\vspace{2cm}

\centerline{\bf Abstract}

\vspace{5mm}

Maximal helicity-violating scattering amplitudes in $N=4$ supersymmetric Yang-Mills theory are dual to Wilson loops on closed null polygons.
We perform their operator product expansion analysis in two-dimensional kinematics in the soft-collinear approximation which
corresponds to the case when some light-cone distances vanish. We construct the expansion in terms of multi-particle ``heavy"-light
operators, where the ``heavy" fields are identified with the Wilson lines defining the OPE channel and the light fields emerge from the curvature 
of the contour. The correlation function of these define the remainder function. We study the dilatation operator for these operators at one 
loop order and find that it corresponds to a non-compact open spin chain. This provides an alternative view on elementary excitations propagating
on the GKP string at weak coupling, which now correspond to particles traveling along an open spin chain. The factorized structure of 
the Wilson loop in the soft limit allows one to represent the two-loop correction to the octagon Wilson loop as a convolution formula and find
the corresponding remainder function.

\end{titlepage}

\setcounter{footnote} 0

\newpage

\pagestyle{plain}
\setcounter{page} 1

Maximal helicity-violating (MHV) scattering amplitudes in maximally supersymmetric Yang-Mill theory are dual to Wilson loops stretched on null polygonal
contours. The latter is built from the on-shell momenta of massless scattered particles. The equivalence of the two objects was hinted at strong coupling
\cite{AldMal07} and then established at weak coupling at one \cite{Kor08,Bra08} and two-loop orders with up to six sides \cite{Kor08} putting the 
conjecture on a firmer foundation. Dual conformal Ward identities \cite{Kor09} allow one to constrain the form of the Wilson loops with an arbitrary 
number of sides. These arise from the known pattern of the dual conformal symmmetry breaking due to ultraviolet divergences of loop integrals from the 
presence of cusps on polygonal contours reproducing the ABDK/BDS ansatz \cite{Ber03}. A conformally invariant remainder function, which 
can only depend on conformal cross-ratios of light-cone distances, enters as an additive component of the full answer for the Wilson loop starting 
from the two-loop order and involving more than five segments on its contour. It is the main unknown in confronting both sides of the duality. Its computation 
in the simplest case of the hexagon serves as a very illustrative example which exhibits inefficiency of the ordinary techniques based on Feynman 
integrals. First, agreement was established numerically between the corresponding six-leg amplitude \cite{Vol08} and Wilson loop \cite{Dru08} 
expressions. The follow-up tedious analyses culminated in analytical results for the latter filling dozens of pages in terms of Goncharov polylogarithms 
\cite{DelDuc10}.  Eventually, the theory of motives \cite{Gon08} allowed one to reduce the entire output to a two-line expression in terms of classical 
polylogarithms \cite{Gon10}. Hence, as this example clearly demonstrates, explicit perturbative computations are extremely complicated, hindering the
unraveling of a simple internal structure of observables and thus techniques that can bypass involved loop analyses, getting to the final answer with 
an efficient shortcut would be highly valuable. A little while ago, a formalism based on the operator product expansion was suggested in Ref.\ \cite{Mal} and 
used to compute the two-loop remained function for polygons in the restricted two-dimensional kinematics \cite{Gai10} as well as the hexagon Wilson loop in 
full $R^{1,3}$ \cite{Gai11} in a very concise manner. Quite recently, this technique was also used to constrain the symbol \cite{Gon08,Gon10} of the three-loop 
hexagon \cite{DixHen11}.

In this note, we provide a complementary view on the operator product expansion by explicitly constructing the first contribution to the series in terms of 
``heavy"-light operators and demonstrating how their renormalization group evolution determines the expectation value of the Wilson loop. This consideration 
also provides an interpretation for Gubser-Klebanov-Polyakov (GKP) excitations \cite{GubKlePol02,AldMal07,Bas10} from the point of view of open spin chains%
\footnote{Unrelated open spin chains had made their appearance in the study of Regge limit of scattering amplitudes in Ref.\ \cite{Lip09}.}. We 
limit our consideration to the two-loop remainder function and thus will not (need to) address loop corrections to the coefficient functions of contributing 
light-cone operators, though this should not represent a problem of principle for the current systematic expansion. 

\begin{figure}[t]
\begin{center}
\mbox{
\begin{picture}(0,160)(230,0)
\put(0,0){\insertfig{16}{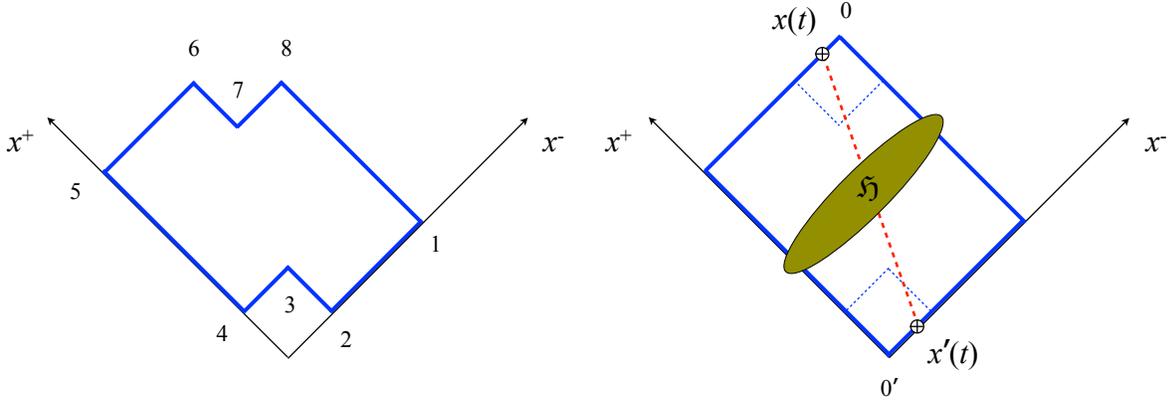}}
\put(327,86){$\mathfrak{H}$}
\end{picture}
}
\end{center}
\caption{ \label{WLope} Null Wilson loop in the $R^{1,1}$ kinematics (left) and leading term contribution to its operator expansion in the soft-collinear
limit $x_{23}^+, \, x_{67}^+ \to 0$.}
\end{figure}

To set up the formalism, we restrict our analysis to the two-dimensional kinematics \cite{Gai10} and thus address the simplest nontrivial case corresponding 
to the octagonal null contour. The object at the center of our analysis is the Wilson loop 
\be
W = \frac{1}{N_c} \vev{\tr \left( W_{[0,1]} \dots W_{[7,8]} W_{[8,1]} \right) }
\, ,
\ee
built from eight light-like segments
\be
W_{[ii+1]} = T \exp \left( i g \int_0^1 dt_i \, \dot{x}(t_i) \cdot A (t_i) \right)
\, ,
\ee
running along particle's momenta as $x(t_i) = x_i - t_i x_{ii+1}$. Below, we will dress the Wilson line segments with the superscripts $\pm$ to designate the 
light-cone direction of the corresponding links on the zig-zag contour (as shown in Fig.\ \ref{WLope}). The Wilson loop expectation value admits the
perturbative expansion
\be
W = \sum_{n = 0} a^n W^{(n)}
\, ,
\ee
in the 't Hooft coupling $a = g_{\scriptscriptstyle \rm YM}^2 N_c/(8 \pi^2)$. Since we will be after the remainder function $W_R$ depending only on the 
conformal cross-ratios, in the current case one can construct just two of these, which we conveniently choose as
\be
u^+ = \frac{x_{32}^+ x_{67}^+}{x_{62}^+ x_{73}^+}
\, , \qquad 
u^- = \frac{x_{87}^- x_{34}^-}{x_{84}^- x_{73}^-}
\, .
\ee

To start with, we will choose a channel for the operator product expansion as in Ref.\ \cite{Mal}, say in the $x^+$ direction. We introduce Wilson lines 
for segments running long the top and bottom side of the loop as follows, $W_{\rm top} = W^-_{[5,6]} W^+_{[6,7]} W^-_{[7,8]}$ and $W_{\rm bot} = W^-_{[1,2]} 
W^+_{[2,3]} W^-_{[3,4]}$, respectively. Then, in the short light-cone distance limit $x_{32}^+ \to 0$ and $x_{67}^+ \to 0$, when the segments adjacent to
these become collinear, the operator product expansion for these two sides can be constructed systematically \cite{CorHas79}, with the leading terms 
having the form
\be
\label{OPEsegments}
\left. W_{\rm top} \right|_{x^+_{67} \to 0} = x_{67}^+ \, \mathcal{F}^-_{[5,0]} W^+_{[0,8]} + \mathcal{O} \left( (x_{67}^+)^2 \right)
\, , \qquad
\left. W_{\rm bot} \right|_{x^+_{32} \to 0} = x_{32}^+ \, W^+_{[4,0']}  \mathcal{F}^-_{[0',1]}  + \mathcal{O} \left( (x_{32}^+)^2 \right)
\, ,
\ee
and being induced by the gluon field strength insertion into the contour,
\be
\mathcal{F}^-_{[5,0]} = i g \int_0^1 dt \, W^-_{[5, t]} \, \dot{x}^- (t) F^{+-} (t) W^-_{[t, 0]} 
\, , \qquad
\mathcal{F}^-_{[0',1]} = i g \int_0^1 dt' \, W^-_{[0', t']} \, \dot{x}^{\prime -}(t) F^{+-} (t') W^-_{[t' ,1]}
\, .
\ee
Here $x (t) = x_{0} - t x_{06}$ and $x' (t') = x_{0'} - t' x_{0'2}$. Notice that the soft-collinear expansion eliminates several cusps of the original Wilson loop and thus
becomes insensitive to the corresponding ultraviolet divergences. However, presently we are interested only in the remainder function which is free from the latter. 
Higher order terms in the series \re{OPEsegments} can be constructed systematically in an analogous fashion and are expressed in terms of multi-field insertions 
along corresponding contours \cite{CorHas79}.

The Wilson loop is determined by the correlation function of the above operators
\be
\left. W \right|_{x^+_{67}, x^+_{32} \to 0} 
= 
x^+_{67} x^+_{32} \vev{\tr \left( W^+_{[0', 5]} \mathcal{F}^-_{[5,0]} W^+_{[0,1]} \mathcal{F}^-_{[1,0']} \right) } 
+ \dots
\, ,
\ee
where the ellipses stands for higher order terms in $x_{32}^+$ and $x_{67}^+$. At leading order in 't Hooft coupling, this correlation
function is simply determined by the one-gluon exchange (see the right panel in Fig.\ \ref{WLope} where one sets $\mathfrak{H} =1$) and 
immediately yields
\ba
\label{LOremainder}
W^{(1)}_R
=
- (\mu \, x^+_{00'})^\varepsilon
\frac{\Gamma (2 - \varepsilon)}{\varepsilon}
\left[
(x_{6 0'}^-)^\varepsilon
-
(x_{0 0'}^-)^\varepsilon
+
(x_{0 2}^-)^\varepsilon
-
(x_{6 2}^-)^\varepsilon
\right] \frac{x^+_{32} x^+_{67}}{(x_{00'}^+)^2}
\simeq
- u^+ \log (1 + u^-)
\, .
\ea
Here the last factor in the dimensionally-regularized expression was replaced (in the final equality) by the invariant cross-ratio $u^+$ to the accuracy 
that we work with since the result is valid for $u^+ \to 0$. In order to restore the full result, we choose another channel for the short light-cone distance
expansion of the Wilson loop, i.e, involving the $x^-$-direction this time. Then, performing the same analysis as done above, one finds analogously 
that $W^{(1)}_R  = - u^- \log(1 + u^+)$ as $x_{34}^-, x_{87}^- \to 0$. Bootstrapping the result, one immediately obtains the exact expression 
$W^{(1)}_R = - \log (1 + u^-) \log(1 + u^+)$ valid for arbitrary values of $u^\pm$ in agreement with \cite{Gai10}.

Beyond the one-gluon exchange, one has to compute corrections to the above correlation function of the light-cone operators $\mathcal{F}^-$.
The calculation is split into the analysis of the one-loop corrections to coefficient functions and evolution kernels. By choosing the light-cone
gauge $A^+ = 0$, one immediately observes that all gauge links on the top and bottom of the loop degenerate to unity, i.e., $W^- = 1$.
Therefore, the only interaction left is of the exchanged $F^{+-}$-field with the Wilson loop segments running long the $x^+$ direction. 

Since we can use the OPE in different channels to reconstruct the full function by means of the bootstrap procedure, we can constrain the
$u^-$-dependence at two loop order by studying the large evolution $x_{00'}^+$-time limit of the correlation function. The leading effect
will arise from the logarithmically enhanced terms. As we can already see from the leading order result the time-evolution is associated
with the dimensionless combination $\mu x_{00'}^+$. Notice that the dependence on the factorization scale $\mu$ will cancel upon accounting 
for the short distance coefficient function which will introduce logarithms involving short light-cone distances entering in dimensionless combinations 
$\mu x_{32}^+$ for the bottom portion of the loop and $\mu x_{67}^+$ for the top one, thus restoring the complete cross-ratio $u^+$ in the sum. 
This is completely analogous to the conventional DGLAP logarithms in the analysis of structure functions of deep-inelastic scattering. Thus, all we 
have to do is to calculate the leading order dilatation operator for the effective light-cone operator
\be
\label{3particleWFW}
\Phi [x_1, x_2, x_3] = h^\dagger (x_1^-) F^{+-} (x^-_2) h(x_3^-)
\, ,
\ee
built from the gluon field strength $F^{+-}$ and two ``infinitely-heavy quark'' fields%
\footnote{The analogy is drawn with heavy-quark effective theory where the heavy quark gets replaced by a Wilson line along its classical path
and interacting with external gauge fields through eikonal vertices. The difference however is that heavy quarks trace time-like trajectories while
we are dealing with light-like ones. However, due to superficial similarity of the two object, we will call the light-like Wilson loop arising in our 
consideration ``heavy" since they will not recoil against massless fields building up light-cone operators similarly to their massive counterparts.} 
$h (x_1) = W^+_{[1,0]}$ and $h (x_3) = W^+_{[3,4]}$, with the paths starting
and ending at the points $x_0 = (x_1^-, L^+)$, $x_3 = (x_3^-, 0)$ and $x_1 = (x_1^-, 0)$, $x_4 = (x_3^-, L^+)$, respectively, and parametrized 
by the corresponding trajectories $x (t) = x_0 - t x_{01}$ and $x (t) = x_3 - t x_{34}$. We set the length of the path $L^+$ equal to the $+$-length of 
the Wilson loop $x_{00'}^+$. This will serve as a regularization parameter of the well-known light-cone singularity of the null Wilson line. We did not display 
the gauge links on the intervals $[1,2]$ and $[2,3]$ since they reduce to unity in the $A^+ = 0$ gauge.

\begin{figure}[t]
\begin{center}
\mbox{
\begin{picture}(0,100)(200,0)
\put(0,0){\insertfig{14}{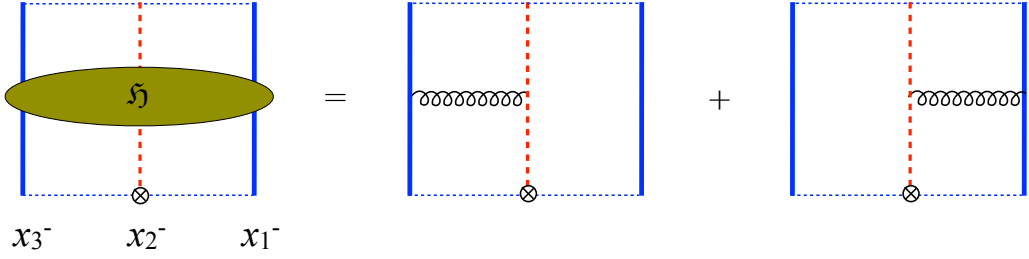}}
\put(51,61.5){$\mathfrak{H}$}
\end{picture}
}
\end{center}
\caption{ \label{DilopOneLoop} One-loop dilatation operator for the ``heavy"-light operator $\Phi$.}
\end{figure}

The dilatation operator can be easily computed from the Feynman graphs shown in Fig.\ \ref{DilopOneLoop}. Making use of the well-known 
coordinate-space perturbative techniques (reviewed, for instance, in Appendix G of Ref.\ \cite{BelRad04}), one immediately finds the following 
result for the pole-part of the one-loop correction to $\Phi$,
\ba
\Phi [x_1, x_2, x_3]
&=&
\left( 1 + \frac{a}{2 \varepsilon} \mathfrak{H}_{123} \right) \Phi_{(0)} [x_1, x_2, x_3]
\ea
where kernel of the dilatation operator has a pair-wise structure at this order
\be
\mathfrak{H}_{123} = \mathfrak{H}_{12}+ \mathfrak{H}_{23}
\ee
with individual Hamiltonians taking the form
\be
\mathfrak{H}_{ii+1} = 
(\mu^2  x_{ii+1}^- L^+)^\varepsilon
\left\{
- \frac{1}{\varepsilon} + \mathfrak{h}_{ii+1}
\right\}
\ee
Notice the presence of the double-pole in $\varepsilon$ which is a familiar consequence of the ultra-violet divergences generated by integrations in 
the vicinity of the cusp points of two adjacent Wilson lines%
\footnote{One of the lines gets degenerated to unity so that the corresponding divergence becomes end-point.} that build up the operator $\Phi$. 
These contributions are accompanied by the one-loop 
cusp anomalous dimension. They will not be of interest for our subsequent consideration since they are exactly captured by the dual
conformal Ward identities \cite{Kor09}. However, the second term in the above expression, which is the focus of our analysis, induces the displacement of 
the light field $F^{+-}$ along the light cone and encodes its renormalization group evolution. Intuitively, only the light degrees of freedom get shifted
since the ``heavy" fields encoded by the Wilson lines do not recoil against the former. The one-loop graphs displayed in Fig.\  \ref{DilopOneLoop}
give
\ba
\label{LOhamiltonians}
\mathfrak{h}_{12} \Phi [x_1, x_2, x_3] 
&=& 
\int_0^1 d \alpha \frac{\bar{\alpha}^{2 s - 1}}{[\alpha]_+} \Phi [x_1, \alpha x_1 + \bar\alpha x_2 , x_3]
\, , \\
\mathfrak{h}_{23} \Phi [x_1, x_2, x_3] 
&=& 
\int_0^1d \alpha \frac{\bar{\alpha}^{2 s - 1}}{[\alpha]_+}  \Phi [x_1, \bar\alpha x_2 +  \alpha x_3 , x_3]
\, ,
\ea
with $\bar\alpha = 1 - \alpha$. Here the calculation was performed for an arbitrary propagating field, --- anticipating follow-up applications to general 
$R^{1,3}$ kinematics and 
supersymmetric Wilson loop, --- which is encoded in the value $s$ of the conformal spin of the excitation%
\footnote{These are gluon field strengths classified with respect to their twist, for the bosonic Wilson loop, or scalars and fermions, for the contour lifted 
to superspace.}. For the case at hand%
\footnote{Notice that the twist-three excitations like $F^{+-}$ are not quasipartonic. As a consequence, the explicit calculation in the light-cone gauge 
$A^+ = 0$ of the $\partial^+ A^- \to \partial^+ A^- $ transitions has to be supplemented by $\partial_+ A_- \to \bit{\partial}_\perp \cdot \bit{A}_\perp$ ones, 
with the latter traded for the $\partial^+ A^-$ contribution by mean of equations of motion. Notice that we also ignored two-to-three transitions which do not contribute
at this order in coupling to the remainder function in the restricted two-dimensional kinematics.}, it is $F^{+-}$ with $s = 1$. In the above dilatation operators we 
introduced a conventional notation 
for the regularized end-point singularity
\baa
\frac{1}{[\alpha]_+} = \frac{1}{\alpha} - \delta (\alpha) \int_0^1 \frac{d \beta}{\beta}
\, .
\eaa
The pairwise Hamiltonian $\mathfrak{h}$ is not $sl_2$ invariant, and commutes with the generators of the light-cone algebra up to boundary terms 
(for $J^-$)
\be
[\mathfrak{h}_{ii+1} , J_{ii+1}^0] = 0  \, \qquad
[\mathfrak{h}_{ii+1} , J_{ii+1}^+] = 0 \, \qquad
[\mathfrak{h}_{ii+1} , J_{ii+1}^-] = x^-_{ii+1} \, .
\ee
The above three-site Hamiltotian $\mathfrak{h}_{123} = \mathfrak{h}_{12} + \mathfrak{h}_{23}$ is also invariant under three-site generators $J^a = \sum_i 
J^a_{i i+1}$ up to the boundary terms,
\be
[\mathfrak{h}_{123} , J^0] = 0  \, \qquad
[\mathfrak{h}_{123} , J^+] = 0 \, \qquad
[\mathfrak{h}_{123} , J^-] = L^- \, ,
\ee
where $L^- = x^-_{13}$ is the length of the operator $\Phi$.

Making use of this result, we can instantly extract the logarithmically enhanced correction to the Wilson loop at two loops. Being after the remainder 
function, we can then promote $\log x_{00'}^+$ to $- \ft12\log u^+$, which is the only conformal-invariant cross ratio involving the ``time''-evolution 
interval. The contribution takes the form of pair-wise kernels of the dilatation operator acting on the ``heavy"-light system that defines the OPE limit of the null 
Wilson loop,
\ba
\label{2loopConvolution}
\left. W \right|_{x^+_{67}, x^+_{32} \to 0} 
&=& 
\frac{a}{2} \, 
x^+_{67} x^+_{32} \log ( \mu \, x_{00'}^+ )
\\
&\times& \vev{\tr \left( \mathfrak{h} [W^+_{[0', 5]} \mathcal{F}^-_{[5,0]}] W_{[0,1]} \mathcal{F}^-_{[1,0']} 
+  W^+_{[0', 5]} \mathcal{F}^-_{[5,0]}  \mathfrak{h} [ W_{[0,1]} \mathcal{F}^-_{[1,0']} ] 
+ 
(x_{0'} \leftrightarrow x_0)
\right) } 
+ \dots
\, . \nonumber
\ea
Here the two-particle operator $\mathfrak{h}$ acts on the ``heavy" fields $W$ and the field strength tensor $F^{+-}$ in $\mathcal{F}^-$ according
to Eqs.\ \re{LOhamiltonians}. In the above equation, we restored the broken conformal symmetry of the evolution kernels by adding a parity-reflected 
contribution $x_{0'} \leftrightarrow x_0$ which restores exact  commutation relations of the $sl{}_2$ algebra. The calculation is straightforward and can 
be simplified making use of the conformal symmetry of the expected result. Making use of the explicit action of the Hamiltonians on pair-wise systems
of fields and choosing the external points as follows $x^-_{0'} = 0, \, x^-_2 = {\rm e}^{- \sigma}, \ x^-_6 = 1, x^-_0 = \infty$ such that $\sigma = \log(1+1/u^-)$, we
obtain the following integral representation of the two-loop remainder function in the soft-collinear limit of the $x^+$-channel, which yields the final answer,
\ba
W^{(2)}_R 
&=& 
- \frac{1}{2}
u^+ \log u^+ \int_0^1 \frac{d \alpha}{[\alpha]_+}
\left\{
\log (1 - \bar\alpha \, {\rm e}^{- \sigma}) + \log (1 - \bar\alpha^{-1} {\rm e}^{- \sigma}) 
\right\}
\\
&=&
- \frac{1}{2} u^+ \log u^+ \log u^- \log(1+ u^-) 
\, .
\ea
Bootstrapping this result by considering the OPE in the cross $x^-$-channel, we reconstruct the two-loop remainder function in the form 
$W^{(2)}_R = -\frac{1}{2} \log u^+ \log(1+ u^+) \log u^- \log(1 + u^-)$. This is indeed the expression obtained by explicit diagrammatic calculations in 
Ref.\ \cite{DelDucDuh10} and confirmed by the OPE techniques in Ref.\ \cite{Mal} (see also \cite{HesKho10}).

The above reformulation of the OPE for the Wilson loop in terms of ``heavy"-light light-cone operators provides a new perspective on the relation of
the Wilson loop to integrable structures. To make a connection to the OPE formalism elaborated in Ref.\ \cite{Mal}, it is instructive to point out a 
relation of the current problem to non-compact open spin chain systems, see Refs.\ \cite{BraDerMan98,Bel99,DerKorMan03}. 
Since the Hilbert space of the pair-wise Hamiltonian does not possess 
lowest/highest weight vectors%
\footnote{This is analogous to the absence of local operator expansion for heavy-light wave functions of the heavy-quark effective theory \cite{Bra03}.}, 
the Algebraic Bethe Ansatz techniques are not applicable to the problems at hand. However, the solution can be 
found making use of the Baxter $\mathfrak{Q} (u)$ operator. Namely, the two-particle Hamiltonian $\mathfrak{h}$ in Eq.\ \re{LOhamiltonians} can be 
deduced from $\mathfrak{Q} (u)$, which obeys the Baxter equation for a two-site open spin chain\footnote{However, only one site is dynamical, i.e, the 
one corresponding to the light field.}
\be
\label{BaxterEq}
(u - i s) \mathfrak{Q} (u - i) = \mathfrak{t}_2 (u) \mathfrak{Q} (u)
\, ,
\ee
where the local integral of motion enters through the function $\mathfrak{t}_2 (u) = u - \mathfrak{p}$ with $\mathfrak{p} = i J^0 = i (x \partial_x + s)$, which 
is the $22$-component of the Lax operator $\mathfrak{L} (u) = u + i {\bit \sigma} \cdot {\bit J}$. Since the equation \re{BaxterEq} is merely a one-term recursion 
relation, it can be easily solved with the result
\be
\mathfrak{Q} (u) = N (u) \frac{\Gamma (i u - i \mathfrak{p})}{\Gamma (i u + s)}
\, ,
\ee
where $N (u)$ is an-$i$ periodic factor, $N (u-i) = N(u)$, which we choose to be one. Then the 
Hamiltonian is constructed from its logarithmic derivative as follows
\ba
\mathfrak{h} = i \left. \frac{d}{du} \right|_{u = - i s} \log \mathfrak{Q} (u) - \psi (2s) + \psi (1) 
&=& - \psi (s - i \mathfrak{p}) + \psi (1)
\nonumber\\
&=& \int_0^1 \frac{d \alpha}{\alpha} [ \bar\alpha^{- i \mathfrak{p} + s -1} - 1]
\, ,
\ea
where we added the anomalous dimension per excitation. This coincides with the hamiltonian computed at leading order of perturbation theory  
\re{LOhamiltonians}. Here the momentum operator $\mathfrak{p}$ possesses continuous eigenvalues and encodes the momentum of the light
field propagating along the chain.

The three-particle Hamiltonian acting on the operator $\Phi$ in Eq.\ \re{3particleWFW}, which is defined on a semi-infinite interval,
\be
\mathfrak{h}_{123}  \Phi [0, x_2 , \infty] 
=
\int_0^1 \frac{d \alpha}{[\bar\alpha]_+} \alpha \Phi [0, \alpha x_2 , \infty] 
+
\int_1^\infty \frac{d \alpha}{[\bar\alpha]_+} \Phi [0, \alpha x_2 , \infty] 
\, ,
\ee
can be easily diagonalized via the Mellin transform
\be
\Phi [0, x_2 , \infty] = \int_C dp \, x_2^{- i p - 1} \widetilde\Phi_p
\, ,
\ee
such that the total Hamiltonian
\be
\mathfrak{h}_{123}  \Phi [0, x_2 , \infty]  =  \int_C dp \, x_2^{- i p - 1} \, \varepsilon(p) \, \widetilde\Phi_p
\, ,
\ee
possesses the eigenspectrum
\be
\varepsilon (p) = \psi (1 - ip) + \psi (1 + ip) - 2 \psi (1)
\, ,
\ee
parametrized by the momentum eigenvalue $p$. This reproduces the energy of a single hole propagating on the background of derivatives 
which describes the elementary GKP excitation at weak coupling \cite{BelGorKor06,Bas10}. Making use of the operator product expansion for 
the null Wilson loop, we can diagonalize the two-loop correction to the remainder function \re{2loopConvolution} by performing the Fourier
transformation in $\sigma$ (or Mellin transformation in $x_2$ as above) of the leading order coefficient function \re{LOremainder} and obtain the 
representation
\be
W^{(2)}_R = - \frac{1}{4} \log (1 + u^+) \log u^+ \int_C \frac{dp}{p \sinh (\pi p)}  {\rm e}^{i p \sigma} \varepsilon (p)
\, ,
\ee
which coincides with the one advocated within the OPE framework of Ref.\ \cite{Mal} for the null polygonal Wilson loop. Here, the contour $C$ runs 
slightly above the real axis. In this equation, we reconstructed the $u^+$-dependence exactly by bootstrapping this expression making use of the cross 
channel as was done above.

The formalism proposed in this note provides an explicit framework for the OPE analysis of the null polygonal Wilson loops. It allows one to clearly 
separate loop effects stemming from coefficient functions and evolution kernels and thus suggests a systematic extension for addressing higher 
order perturbative corrections. Our consideration also proposes a novel interpretation of the GKP excitations from the point of view of a 
non-compact open spin chain rather than a closed Heisenberg magnet picture involved in the analysis of the large-spin asymptotics of anomalous 
dimensions of twist operators. In the latter, the GKP modes emerge as holes propagating on the background of derivatives building up Wilson operators. 

Recent applications of the operator product technique to the supersymmetric case \cite{VieSev11} do not heavily rely on the pertinent super Wilson 
loop \cite{MaSki10,Car10} and thus do not directly elucidate its role in the relation to non-MHV amplitudes which currently is not defined beyond
tree approximation due to lack of a supersymmetry preserving regularization scheme required for its robust analysis \cite{BelKor11}. The method
employed in Ref.\ \cite{VieSev11} is not particularly sensitive to the specifics of the coupling of GKP excitations to the super-contour of the loop. We 
hope that the technique we advocated in this paper will allow one to probe the connection between the scattering amplitudes and super Wilson loops
beyond MHV approximation in a straightforward fashion. It would also be interesting to study higher-loop corrections to the open-spin chain hamiltonian 
and establish a relation to the framework of factorized scattering of hole excitations on the large-spin closed noncompact spin chains. 

\vspace{0.5cm}

We would like to thank Benjamin Basso for interest in the project at its very early stage and Gregory Korchemsky for reading the manuscript and 
instructive remarks. This work was supported by the National Science Foundation under grants PHY-0757394 and PHY-1068286.


\end{document}